\documentclass[12pt]{article}

\usepackage{epsfig}
\begin{document}

{\bf \large Occurrence of Hysteresis like behavior of resistance of ${\rm
Sb_2Te_3}$ film in heating-cooling cycle.}

\vskip 0.25cm
\begin{center}
{\sl P. Arun \footnote{arunp92@physics.du.ac.in}, Pankaj Tyagi \& 
A. G. Vedeshwar \footnote{agni@physics.du.ac.in}$^*$\\
Department of Physics \& Astrophysics, 
University of Delhi, Delhi - 110 007,\\ 
India}

\vskip 0.25cm
{\bf Abstract}
\end{center}
Experimental observations of a peculiar behavior observed on heating and
cooling ${\rm Sb_2Te_3}$ films at different heating and cooling rate are 
detailed. The film regained its original  resistance, forming a closed loop, 
on the completion of the heating-cooling cycle which was reproducible  
for identical conditions of heating and cooling. The area enclosed by the
loop was found to depend on (i) the thickness of the film, (ii) the heating 
rate, (iii) the maximum temperature to which film was heated and (iv) the 
cooling rate. The observations are explained on basis of model which 
considers the film to be a resultant of parallel resistances. The film's 
finite thermal conductivity gives rise to a temperature gradient along the 
thickness of the film, due to this and the temperature coefficient of
resistance, the parallel combination of resistance changes with temperature. 
Difference in heating and cooling rates give different temperature gradient, 
which explains the observed hysteresis.\\


\vfil\eject

\section{Introduction}
The electrical conductivity measurement is quite an important
characterization technique for thin films and are routinely carried out for 
various materials ranging from metals to high resistive semiconductors. The 
temperature dependence of resistivity yields information about intrinsic band 
gap of  material, activation energy for conduction in polycrystalline films 
(grain boundary barrier height), impurity activation energy etc. In most of 
these measurements data are taken either in heating or cooling direction of 
temperature variation. If the system is homogeneous and the rate of 
change of temperature is constant, no hysteresis can be expected in 
heating-cooling cycle. However, considerable hysteresis has been observed 
when an amorphous film was heated above crystalline transition temperature
and cooled back which can be understood as due to structural changes
\cite{1, 2, 3}. Such hysteresis have been observed in Bi films even without 
structural changes \cite{4}. A hysteresis behavior was also observed in Pd 
film which was ascribed to absorption and desorption of hydrogen during 
heating-cooling cycle \cite{5}. Therefore, it motivated us to investigate
more thoroughly the temperature dependence of electrical resistance in heating-
cooling cycles for ${\rm Sb_2Te_3}$ film in vacuum. We report here the 
interesting observations of hysteretic behavior in repeated cycles of heating-
cooling. The results can be explained by the parallel resistor model \cite{6arun}. 

\section{Experimental Details}

Thin films of ${\rm Sb_2Te_3}$ were grown by thermal evaporation using a 
molybdenum boat on microscope slides glass substrates at room temperature 
at a vacuum better than ${\rm 10^{-6}}$ Torr. The starting material was
99.99\% pure stoichiometric ingot supplied by Aldrich (USA). Films were grown 
on the pre-deposited indium contacts using a mask for the four probe 
configuration of resistivity measurements. The electrical constacts were
confirmed to be ohmic for all the samples studied. The temperature of the film 
was measured by a thermocouple placed very near to the sample on the substrate. 
The structure, chemical composition and morphology of the films were determined 
by X-ray diffraction (Philip PW1840 X-ray diffractometer), photo-electron 
spectroscopy (Shimadzu's ESCA 750) and scanning electron microscope (JOEL-840)
respectively. The film thickness was monitored during it's growth by quartz 
crystal thickness monitor and was subsequently confirmed by Dektak IIA surface 
profiler, which uses the method of a mechanical stylus movement on the surface. 
The movement of the stylus across the edge of the film determines the step 
height or the film thickness. All the films were found to be stoichiometric and 
micro-crystalline after they 
were allowed to age in vacuum for few weeks \cite{6}. Various films of 
thickness between 100 to 500nm were used during the experiment. All 
the heating-cooling cycles for resistivity measurements were carried out in
vacuum of about ${\rm 10^{-6}}$ Torr. This was done to prevent any oxidation of 
the films while heating, which would result in an irreversible change in the
film resistance. The samples were heated by placing them on a copper block 
which is heated by a heating coil embedded in it. 

The heating rate was varied by the voltage applied to the heating coil. The
temperature and the resistance of the film was measured as a function of time. 
The heater was switched off for cooling the film. The cooling of the film from 
its elevated temperature was very slow as the measurements were done in vacuum. 
Therefore, while some control of the heating rate was maintained by controlling
the voltage applied to the heating coil, the cooling rate was essentially
determined by the maximum temperature at which the heater was switched off.
The heating and cooling rates hence, were largely different. 

Figure 1 shows the variation of temperature with time for both heating and
cooling cycles. The temperature increases with time as indicated in segment 
"AB". The heater was switched off on attaining the pre-determined
temperature. Point "B" corresponds to the instant the heater was switched off. 
This is followed by a period of constant temperature "BC". The onset of cooling 
process starts at point "C". The cooling of the films in vacuum could be
either by radiation losses or by conduction through the substrate side or
both. After the heater is switched off, 
since the process is in vacuum, the substrate temperature remains constant for 
an appreciably long time before it starts falling. From figure 1, we find
that the fall in substrate temperature commences 200sec later ater the 
heater is switched off. Beyond point `C' the cooling process starts. However, 
as explained earlier, since the cooling process takes place in vacuum, it is 
very slow, hence the figure does not show the attainment of room temperature 
on cooling. Data of region "AB" is best fitted by the equation
\begin{eqnarray}
T_{sub}(t) = T_{max}(1-e^{-Qt}) + T_{rt}
\end{eqnarray}
The equation shows saturation of temperature with time when the heating due
to the heater is compensated by cooling. ${\rm T_{rt}}$ is the room
temperature. The magnitude of ${\rm T_{max}}$
and "Q" are related in a complex manner to the heater voltage, cooling
mechanism etc. An estimate of these factors are made by fitting equation (1)
to the experimental data shown in figure 1. The value of coefficients ${\rm
T_{max}}$, "Q" and ${\rm T_{rt}}$ were estimated as 360$^o$C, 0.00039${\rm
sec^{-1}}$ and 14.5$^o$C respectively. From equation (1), the rate of
heating during the heating cycle is given by
\begin{eqnarray}
{dT_{sub} \over dt} = T_{max}Qe^{-Qt}
\end{eqnarray}
Similarly, the cooling rate can be found by fitting data of the cooling region
by the equation
\begin{eqnarray}
T_{sub}(t) = T_{final}e^{-St} + T_{rt}
\end{eqnarray}
where ${\rm T_{final}}$ is the final temperature attained before the heater is 
switched off. In the present work ${\rm T_{final} < T_{max}}$. The cooling
rate is governed by "S" and in our case since cooling was occurring naturally
in vacuum (${\rm 10^{-6}}$Torr), "S" was very low $\sim$0.00021${\rm
sec^{-1}}$.

\section{Results and Discussions}
The resistance of as grown film was found to be high and decreases with time
saturating in a few weeks \cite{6}. No further change in resistance was 
observed with time thereafter. The as grown amorphous films were found to age 
into a polycrystalline state, with peak positions matching those given 
in ASTM card 15-874. The film resistance follows a hysteresis path with 
heating-cooling cycle which was reproducible under identical conditions.
Reproducibility was tested as many as 20 times in some samples. The films were 
heated to various temperatures below ${\rm 250^oC}$ (${\rm
T_{final}}$) in vacuum. It was ensured no change in material properties like
structural, compositional or morphological took place by repeated
characterizations after each cycle or few cycles. This was quite crucial in
ruling out the contributions from these parametric changes to irreversibility
or hysteretic behavior of resistance with temperature.  Figure 2 shows the 
X-ray diffractograms of a 210nm, film after repeated heating-cooling cycles. 
As can be seen from the figure, even 
after the seventh cycle of heating and cooling, there was no change in the 
crystal structure or any improvement in the grain size etc. Physical and 
chemical changes  hence can be ruled out as to be occurring due to the 
heating-cooling cycles. For further comphrehansive understanding of the
hysteretic behavior, we have carried out experiments varying only one
parameter and keeping other same as described below.
 
\subsection{ Film thickness dependence} Films of different thickness were 
heated at the same heating rate to reach identical maximum temperature 
(${\rm T_{final}}$). Figure 3 shows the hysteresis paths taken during the 
heating and cooling processes for various film thicknesses. We have marked three 
different regions as "AB", "BC" and "CD" on the hysteresis loop. The resistance 
decreases with increasing temperature (region "AB") indicating the semiconducting 
nature of the ${\rm Sb_2Te_3}$ film, which is a p type narrow band gap material 
\cite{7}. Point "B" marks the point where the heater is switched off. Even though 
the heater was switched off, the temperature of the sample does not 
decrease immediately as the measurements were done in vacuum (see figure 1). 
However, the resistance continues to fall till point "C" at the almost
constant temperature within some span of time. Thus, variation in
resistance in region "BC" is with time. Hence, in figure 4, we show the
variation of resistance with time. As can be seen, the heater is switched
off at 'B' and the resistance of the sample continues to decrease with time till 
the onset of the cooling process marked by point "C". It may be noted that
the measured temperature in the region 'BC' was constant as shown in figure
3 and the decrease in resistance looks to be very steep in this region.
Figure 4 depicts the same fact, that is, increasing resistance with decreasing 
temperature in cooling process beyond 'C' as seen in fig 3. The increase in
resistance during cooling in fig 3 (between 'CD') is almost parallel to the
decrease in segment "AB". At point "D", the film reaches room temperature and 
the film resistance goes back to point 
"A", very slowly, over a long time ($\sim$ 8-9 hours). The regions "BC" and 
"DA" are quite puzzling, where the temperature is constant and resistance is 
varying with time. It is evident that films of different thickness films enclose 
different area under the loop. Figure 5 shows the variation of the area 
enclosed under the loop for various thickness. The graph was plotted using
the measured area enclosed by various loops of figure 3. 

\subsection{ Dependence of final temperature \& cooling rate} Samples of 
identical thickness 
were heated with the same heating rate but to different final temperatures by 
switching off the heater at different temperatures (${\rm T_{final}}$). The 
segment "AB" in such cases coincided. However, the length of the segment
"BC" varied. Figure 6 compares two cases (for clarity only two are shown) 
where the same film was heated at the same heating rate, but to two different 
${\rm T_{final}}$. The cooling rate was not controllable in our present 
experiment, as it was allowed for natural cooling in vacuum. It is clear that 
the cooling rate strongly depends on ${\rm T_{final}}$. From the above two cases 
it is implied that the area under the loop is also dependent on the cooling rate 
of the films. The explicit dependence of cooling rate could also be studied
with a convenient and controllable cooling arrangement which is not possible
in the present study.

\subsection{ Dependence of heating rate} To understand the effect of heating 
rate on the area enclosed by the loop, films of same thickness were heated at 
different heating rates, as shown in figure 7. From equation 2, the heating rate 
also varies with time. It's magnitude depends on ${\rm T_{max}}$ and Q, both of 
which can be controlled or selected by the voltage applied to the heater. 
It is evident from figure 7, dR/dT is greater for lower heating rate. The slope 
is larger for lower rate of heating, resulting in smaller area enclosed. Hence 
the area enclosed increases with increasing heating rate. In other words the 
rate of change of resistance, or the thermal coefficient of resistance depends 
on the heating rate. 

In summary the area under the hysteresis loop hence, was found mainly to depend 
on the following parameters of the experiment:\\
(a)  the film thickness\\
(b)  the heating rate \\
(c)  the final temperature (${\rm T_{final}}$) that the sample was heated to 
and\\
(d)  the rate of cooling \\
\par Most of these observations can be explained qualitatively suing the
theory in \cite{6arun}. However, we briefly outline the theory and the model 
below.

\section{Theory}

We first calculate the temperature profile along the film thickness which
need not be uniform especially during heating/ cooling process in a dynamic
or transient measurement. Then the calculation of the total resistance of
the film as a function of temperature can simply be carried out by
integrating across the film thickness. This should be the key in explaining
the observed hysteresis behavior. As described earlier the film is kept on 
copper block being heated. Heating proceeds from the substrate 
side. Therefore, the temperature varies along the film thickness with time which 
is essentially a one dimensional problem of heating conduction across the film
thickness. The variation of temperature with time and spatial co-ordinates is 
given by\cite{8}
\begin{eqnarray}
c_v{\partial T \over \partial t} = \lambda {\partial^2 T \over \partial x^2}
\end{eqnarray}
where ${\lambda}$ is the thermal conductivity of the film and ${\rm c_v}$ is 
the specific heat of the film. A solution of this partial differential 
equation depends on the initial and boundary conditions of the problem. 
Depending on the initial and boundary conditions solution would be 
different\cite{9}. 
For the given experimental conditions the variation of temperature with
spatial and time co-ordinates is given by [6]. The variation in temperature
along the film thickness with time is given as
\begin{eqnarray}
T(x,t)= T_{sub} - (T_{sub}- T_{sur})sin\left ( \pi x \over 2 d \right ) 
e^{-{\pi^2 D t \over 4 d^2}} 
\end{eqnarray}
where D is the thermal diffusivity (${\rm \lambda / c_v }$) and d is the
film thickness. The temperature profile across the film thickness can be
calculated using equation 7. At the starting of heating, ${\rm T_{sur}}$ in
the equation can simply be taken as room temperature with ${\rm T_{sub}}$
slightly hotter by few degrees. Every time a new resulting ${\rm T_{sur}}$
is used along with the incremented ${\rm T_{sub}}$. Thus, the profile can be
calculated numerically. ${\rm T_{sub}}$ serves as the heat source.
Obviously, the difference between ${\rm T_{sur}}$ and ${\rm T_{sub}}$ would
increase with decreasing D or ${\rm \lambda}$ of the given material
exhibiting a quite non-uniform temperature distribution along the film
thickness at the given instant of time. The time for reaching equilibrium or
uniform distribution is also inversely proportional to ${\rm \lambda}$ of
the material because ${\rm c_v}$ does not vary much from material to
material at high temperatures.

The film can be thought of as a stack of numerous infinitesimal identical thin 
layersof same thickness. All the layers acting as resistive elements with the
net resistance of the film as the resistance in parallel combination of
these layers. Since the layers are identical, at room temperature all 
of them have equal value. However, due to the metallic/ semiconducting nature  
of the film, the resistance of these layers vary with temperature. For
simplicity, the variation of resistance with temperature is taken linear as
\begin{eqnarray}
R_{layer} = R_o(1+ \alpha T)
\end{eqnarray}
where ${\rm \alpha}$ and ${\rm R_o}$ are the temperature coefficient of 
resistance (TCR) and the resistance of the identical layers respectively. For 
the case ${\rm T=0^oC}$, the films resistance would be given as  
\begin{eqnarray}
{1 \over R_{film}} = \sum_{i=1}^{i=n} {1 \over R_o} = {n \over R_o} 
\end{eqnarray}
The TCR is positive for metal while it is negative for semiconductors. Since, 
spatial distribution of temperature along thickness was calculated for various 
substrate temperatures at various instant, the films resistance can be trivially 
calculated as a function of substrate temperature and time.

We have calculated the film resistance as a function of temperature as
described above using eqn(5-7). We have taken 100nm thick film as a stack of
10 identical layers in parallel combination with each layer's resistance of
${\rm 170K\Omega}$ at room temperature and ${\rm \alpha=-0.8 \times
10^{-3o}C^{-1}}$. These numerical values are taken from our previous study
on ${\rm Sb_2Te_3}$ films [11]. The results are plotted in fig 8 for varying
diffusivity or mainly the thermal conductivity. The visual examination of
fig 8 reveals a peaking behavior of hysteresis loop area with thermal
conductivity of the film. We have, therefore, plotted the hysteresis loop
area exclusively as a function of diffusivity in fig 9. The loop area shows
a maximum at intermediate diffusivity. This may look very surprising on the
onset. However, there is a striking similarity between fig 5 and fig 9, that
is dependence of loop area on film thickness and thermal conductivity. Fig 9
is the direct consequence of varying thermal conductivity as calculated by
the above model resulting from the temperature profile across the film
thickness. The film resistance, the parallel combination of identical
resistive layers would crucially depend on the temperature profile across
the thickness. The results could be almost similar for very low thermal
conductivity and very high thermal conductivity due to nearly uniform
temperature profile. Therefore, an increased or enhanced loop area for
moderate thermal conductivity seems to be quite reasonable arising due to
quite non-uniform temperature profile across film thickness. 

The similarity between the behavior of loop area with film thickness and
thermal conductivity may also be expected because many physical properties
like thermal conductivity show thickness dependence [12]. The exact
dependence may vary from material to material. In the present study, ${\rm
Sb_2Te_3}$ films are semi-metallic and shows a linearly inverse relation of
resistance with film thickness [11]. in the range of fig 5. Since films are
semi-metallic, by Wiedemann-Franz law, we can see that thermal conductivity
varies linearly with film thickness. Therefore, the experimental result of
thickness dependence of loop area shows an analogous behavior to that
predicted by the calculated dependence of loop area on thermal conductivity.
However, the resistivity od ${\rm Sb_2Te_3}$ films is slightly lower larger
due to its polycrystalline nature. Still we feel it is within the
applicability of Wiedemann-Franz law. Also, the thermal conductivity used in
the model calculation is the total of lattice and electron contributions.
Further, we have observed hysteresis loops even in the amorphous films of
${\rm Bi_2Te_3}$ as shown in fig 10. Similarly, the polycrystalline InSb
films also show hysteresis. The present model explains quite well
qualitatively the features of the hysteresis behavior. At present we have
not fitted the experimental data with the model due to thenon-availability
of few material parameters required. Alternatively one can estimate thermal
conductivity of the film across its thickness by fitting the experimental
data and the model. It would be same along parallel and perpendicular
directions of the film for isotropic materials and differ for anisotropic
materials. We are pursuing few other different materials in this direction
along with quantitative analysis and fitting of experimental data for the
determination of thermal conductivity. However, the detailed analysis will
be the subject for future publication.

\section{Conclusions}
The electrical studies of thin films are usually done by heating the sample
and measuring resistance/ resistivity with temperature. Though, the
measurements are to be done after the film has attained a steady
temperature, usually the measurement is done as the film is being heated or
cooled. As discussed in the article, if the film has a finite thermal
conductivity, one essentially is making
measurement in non-equilibrium conditions. Thus, parameters like TCR etc.
computed is not only material dependent but depends on conditions of the
experiment, e.g. the rate of heating or cooling. It is essentially due to
this non-equilibrium measurement that leads to a loop like formation due to
the heating-cooling cycle. Where the area enclosed by the loop depends on
the films' thermal conductivity, rate of heating and cooling. This method
may be developed to index the film's diffusivity.

\vfil\eject

\pagebreak

\begin{figure}
\begin{center}
\epsfig{file=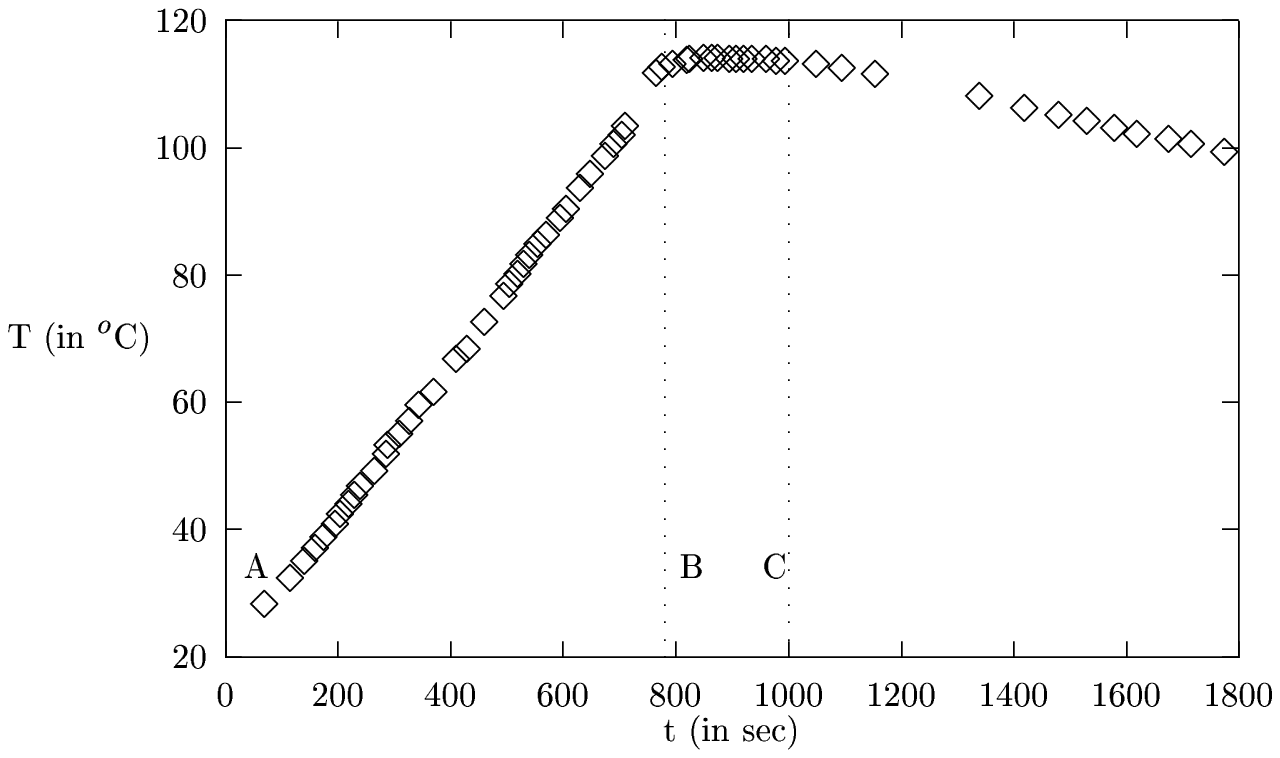,width=5 in}[tbh]
\caption{ Variation of temperature with time during the heating-cooling
cycle. Region 'AB' and 'BC' are indicated.} 
\label{fig:1}
\end{center}
\end{figure}

\begin{figure}
\begin{center}
\epsfig{file=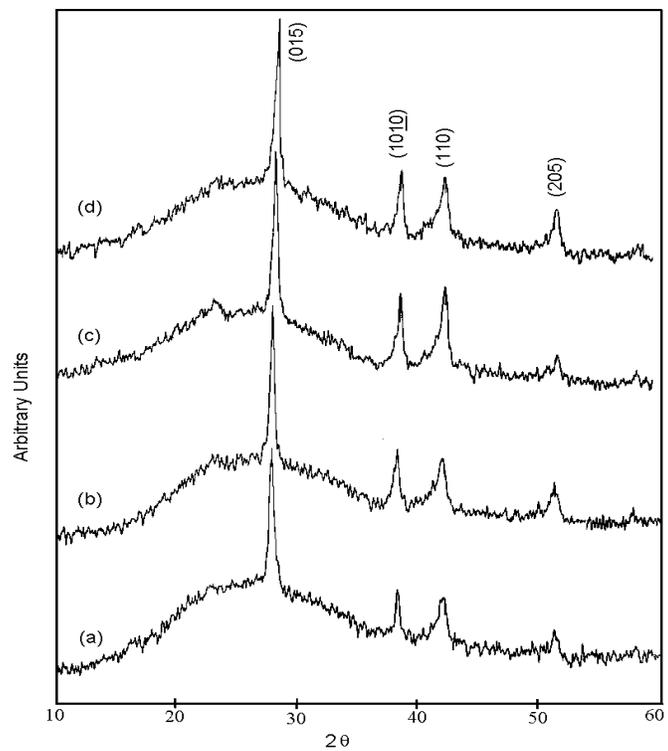,height=5in, width=3.5 in}
\caption{X-ray diffractograms of ${\rm Sb_2Te_3}$ film after (a) one, (b)
two, (c) six and (d) seven heating-cooling cycles.} 
\label{fig:2}
\end{center}
\end{figure}

\begin{figure}
\begin{center}
\epsfig{file=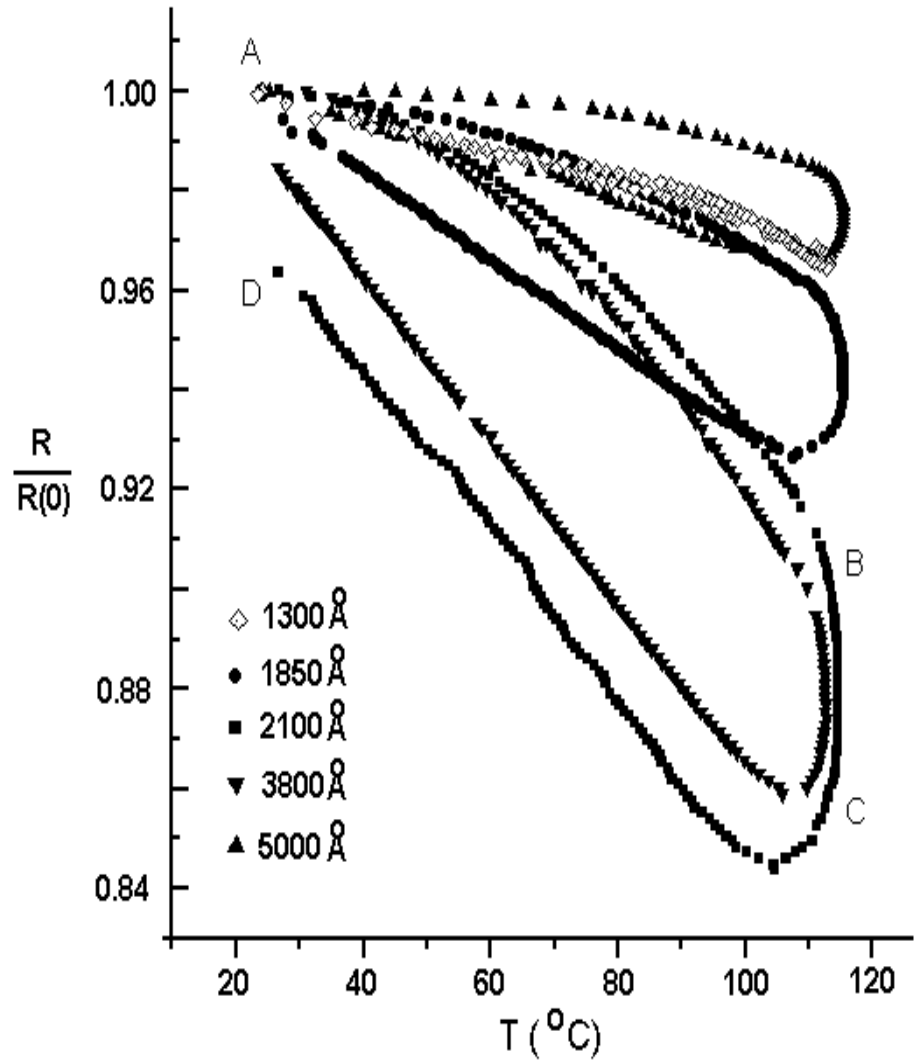,height=6in, width=5 in}
\caption{Variation of resistance with temperature during the heating and
cooling cycles for films of thickness (a) 1300\AA\, (b) 1800\AA\, (c)
2100\AA\, (d) 3800\AA\, and (e) 5000\AA. The maximum temperature to which
the films were heated and the heating rate was same in both cases. Regions
"AB", "BC" and "CD" are indicated.}
\label{fig:3}
\end{center}
\end{figure}

\begin{figure}
\epsfig{file=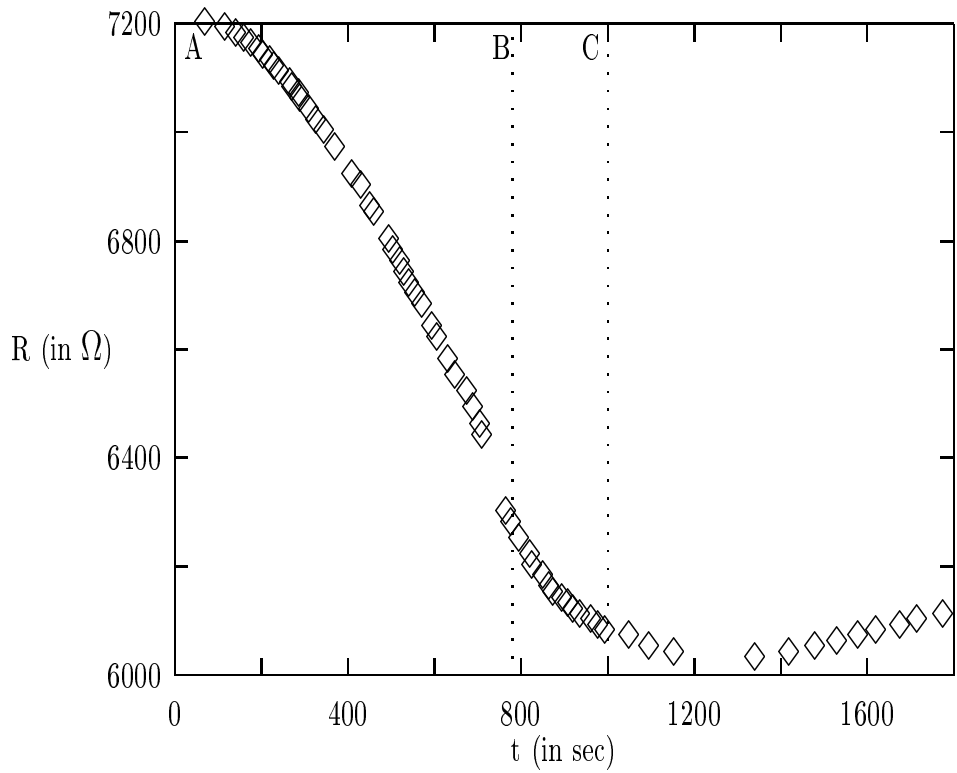,width=5in,height=4in}
\caption{Variation of resistance with time during the heating-cooling cycle.
Regions "AB" and "BC" are indicated.}
\label{fig:4}
\end{figure} 

\begin{figure}
\begin{center}
\epsfig{file=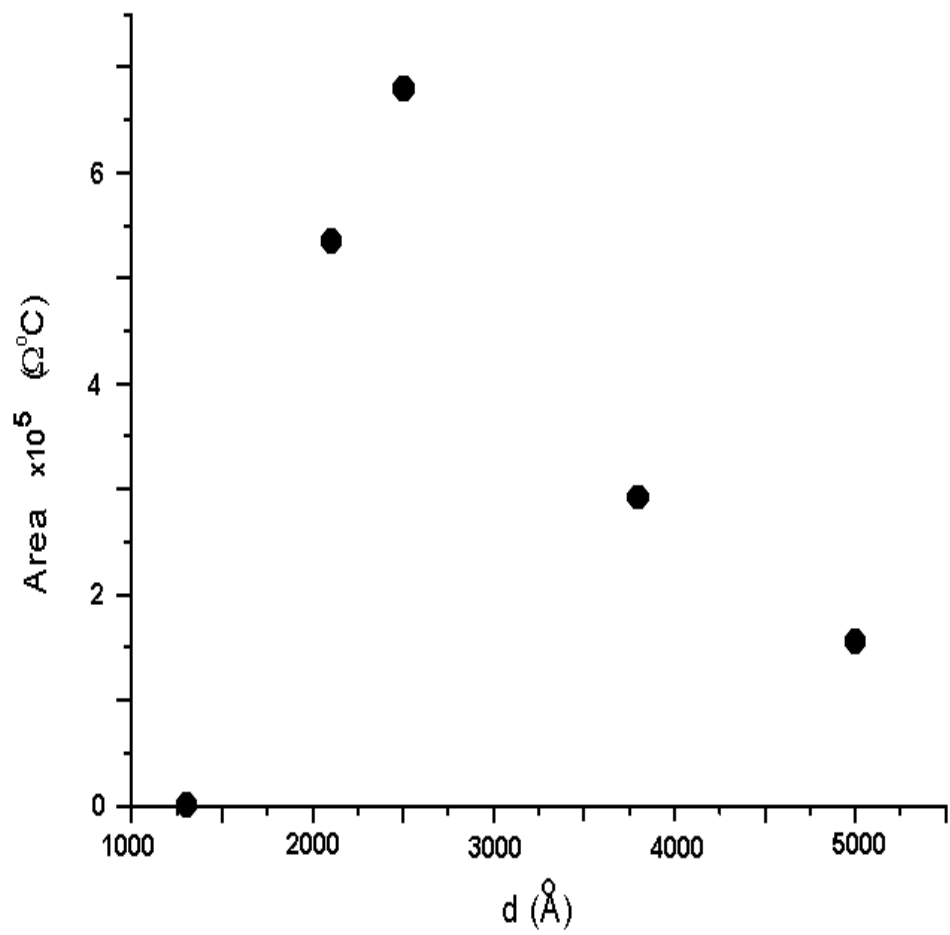,height=5in, width=5.25 in}
\caption{Area enclosed under the hysteresis loop of different film
thickness.} 
\label{fig:5}
\end{center}
\end{figure}

\begin{figure}
\begin{center}
\epsfig{file=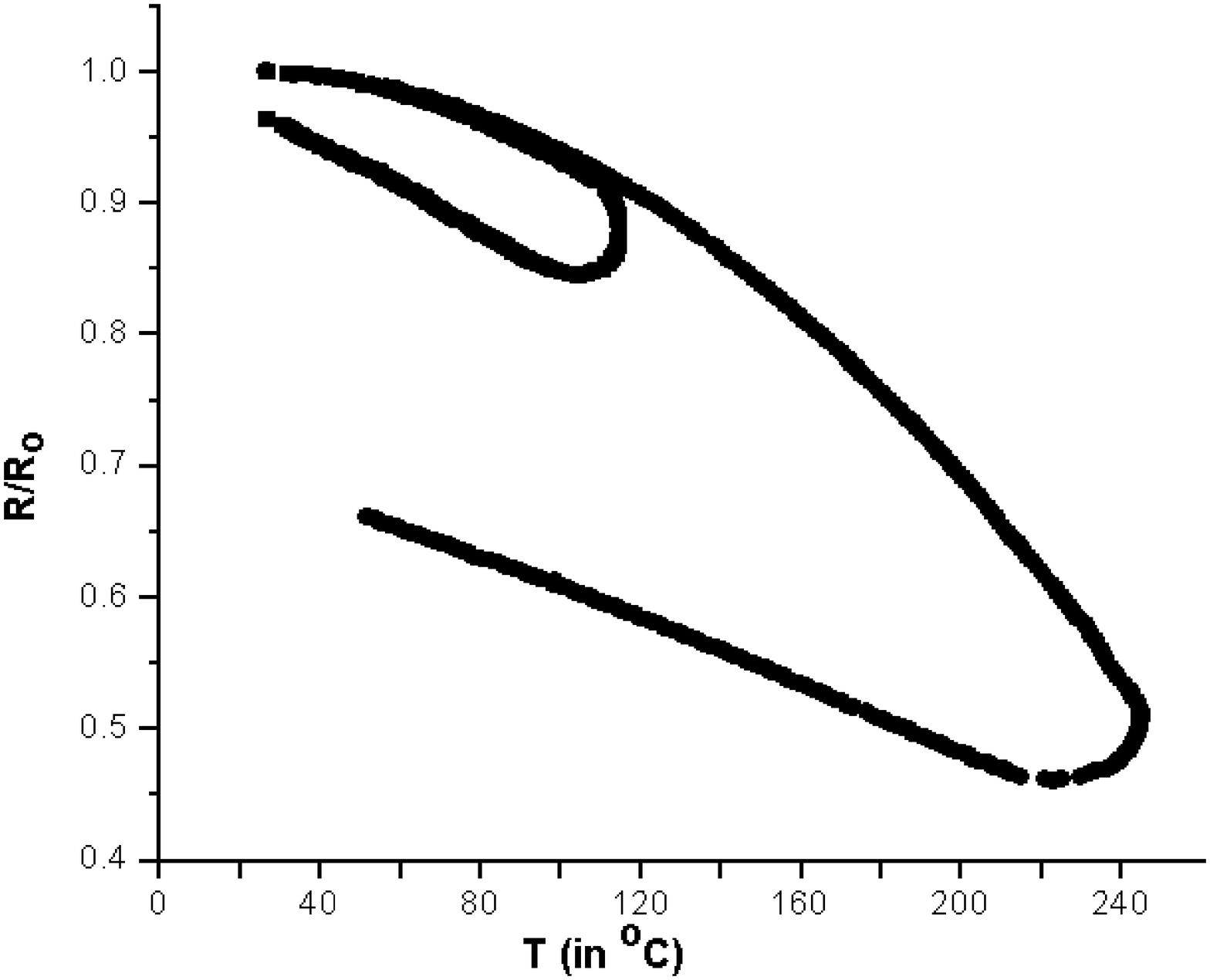,height=5in, width=5.25 in}
\caption{ A ${\rm Sb_2Te_3}$ film of thickness 2100\AA\, was heated to two
different ${\rm T_{final}}$. 110$^o$C and 250$^o$C at same heating rate.}
\label{fig:6}
\end{center}
\end{figure}

\begin{figure}
\begin{center}
\epsfig{file=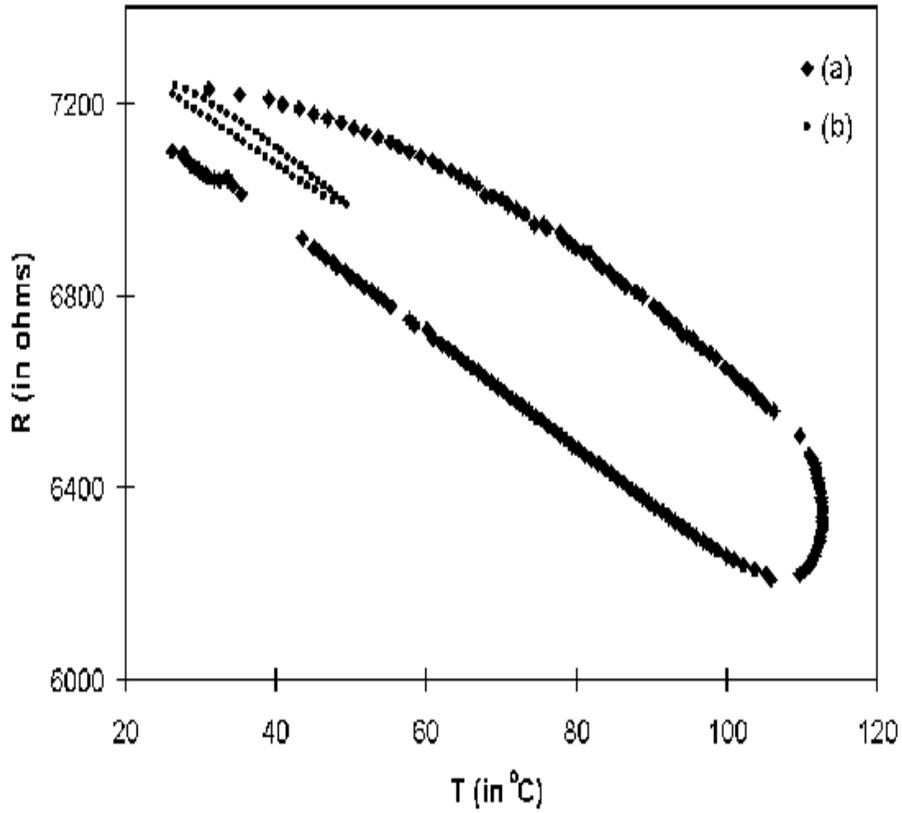,height=4.5in, width=5 in}
\caption{Variation of resistance in the heating and cooling cycle for a film
of thickness 3800\AA. The heating rate was kept different by supplying the
heater different voltage. Parameter "Q" of equation (1) used were (a) ${\rm
4.82 \times 10^{-4}sec^{-1}}$ and (b) ${\rm 1 \times 10^{-5} sec^{-1}}$.}
\label{fig:7}
\end{center}
\end{figure}

\begin{figure}
\begin{center}
\epsfig{file=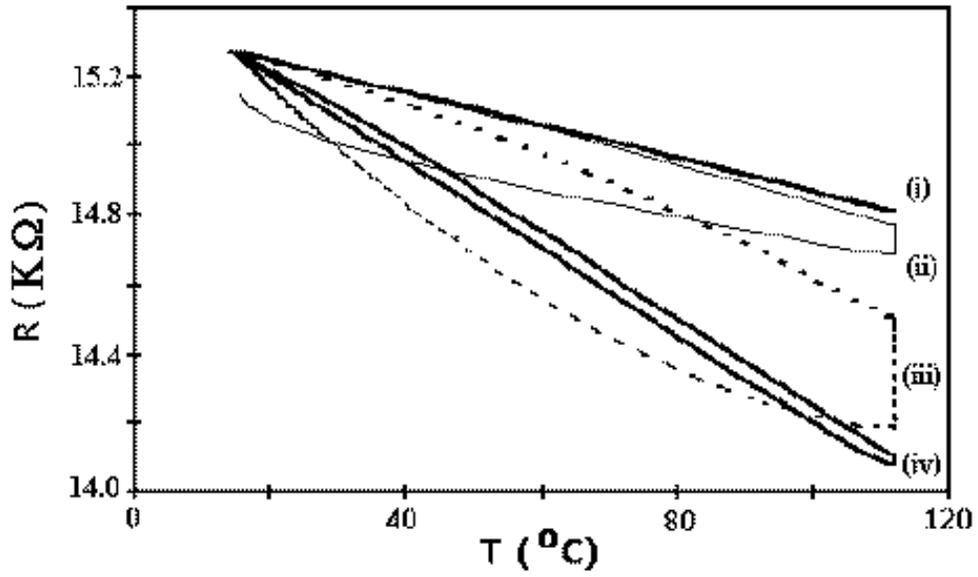,width=5.25 in}
\caption{  
Hysteresis loops formed in film resistance with the heating-cooling cycle.
The calculations were done for film thickness of 1000\AA\, and diffusivity (i)
${5 \times 10^{-3} \AA^2/sec}$, (ii) ${50 \AA^2/sec}$, (iii) 
${5 \times 10^2 \AA^2/sec}$ and (iv) ${5 \times 10^3 \AA^2/sec}$.}
\label{fig:12}
\end{center}
\end{figure}

\begin{figure}
\begin{center}
\epsfig{file=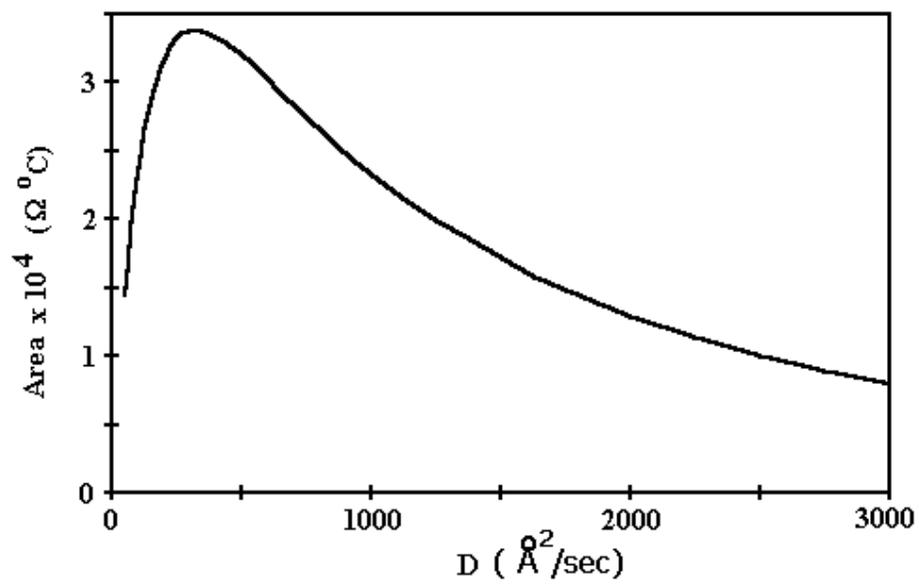,width=5 in}
\caption{ The variation in the area enclosed by loops formed during the
resistance variation with temperature during heating-cooling cycles. The
variation is due to the difference in the films diffusitivity. }
\label{fig:13}
\end{center}
\end{figure}

\begin{figure}
\begin{center}
\epsfig{file=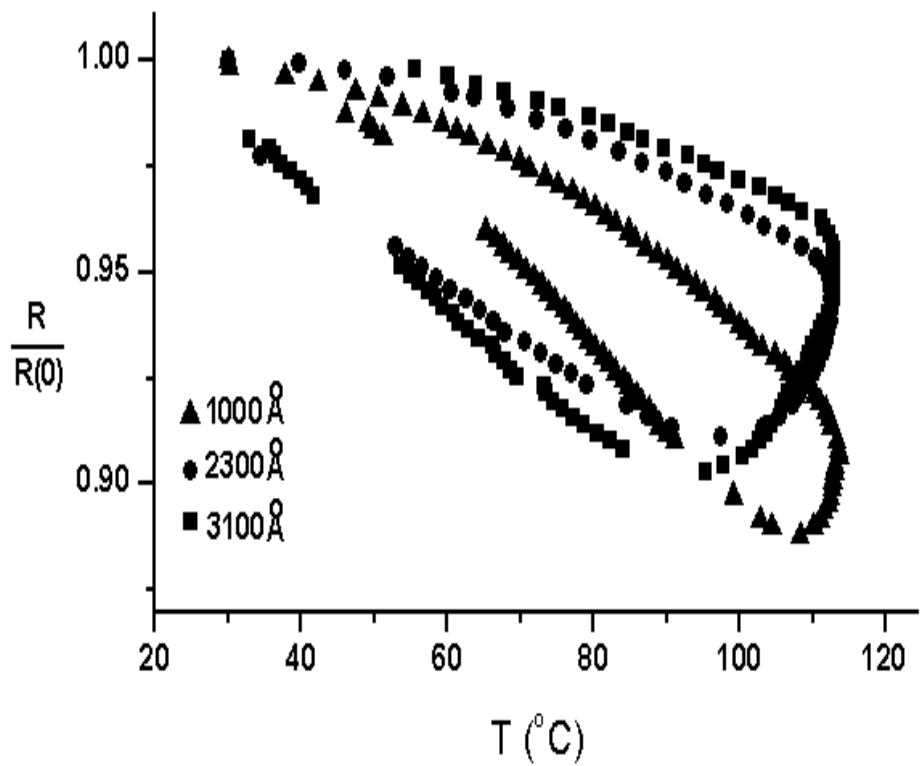,height=4.5 in, width=5 in}
\caption{Variation of resistance with time during the heating-cooling cycle
in ${\rm Bi_2Te_3}$.}
\label{fig:bi2te3}
\end{center}
\end{figure}

\end{document}